\documentclass[reprint,amsmath,amssymb,aps,showpacs]{revtex4-1}  
\pdfoutput=1

\usepackage{graphicx}
\usepackage{dcolumn}
\usepackage{bm}  
  
\begin{document}

\title{   } 

\author{ }

\affiliation{$^{\star}$ Instituto de F\'{\i}sica, Universidade Federal do Rio de Janeiro, 
Av. Athos da Silveira Ramos, 149,
Cidade Universit\'aria, 21941-972, Rio de Janeiro, RJ, Brazil}
\affiliation{$^{\ddag}$ }




{\bf Comment on ``Universal and accessible entropy estimation using a compression algorithm"}\\

In a recent Letter \cite{Avinery} a framework for estimating entropy 
was introduced and applied to one-dimensional and two-dimensional systems.
In this Comment we show that the method is not well suited for estimating  entropy in 
bidimensional systems presenting long-range correlations.

In Ref. \cite{Avinery}, entropy is evaluated following this  scheme:
1) discretize the considered configurations
2) store them in a 1D file 
3) measure the compressed file size with a  lossless compression algorithm    
4) estimate the incompressibility 
and map it to the asymptotic entropy $S_A$.
Here we test this scheme for binary variables,
and we do not consider the 
complication produced by the coarse-graining of continuous variables.

In general, 
lossless compression algorithms are known to present slow entropy convergence
and alternative  
more efficient methods are 
used \cite{Grassberger96}. 
There is a rich literature studying 1D systems
but the analyses of multidimensional patterns are very few, 
and they are expected to present new features compared to the 1D case.
Some methods are known to apply to patterns of arbitrary dimension
(e.g. block entropies).
In contrast, mapping multidimensional patterns to a one-dimensional sequence 
can be path-dependent, loses bidimensional correlations, and can even produce 
spurious long-range correlations 
\cite{Grassberger86}. The use of a locality-preserving curve, like Hilbert's one,
does not guarantee to solve these difficulties.

Systems which present long-range correlations can be used to test these effects.
In the original Letter, long-range correlations are matched only 
in the 2D NN Ising model near the critical point. 
In that region, the estimated $S_A$ displays a relative error between 5\% and 10\%.
Unfortunately, 
this Ising model is a particular exception where the reduction to a 1D string 
is known to not affect
the statistics that determine 
entropy estimation \cite{Ising1D}.
For this reason, this system is not a 
good benchmark for testing entropy estimation in 2D patterns with long-range correlations. 
For testing the Avinery's framework in this situation, here we consider a set of 68  
different patterns obtained from built form maps of cities around the world \cite{nos2}.
These binary matrices of size $1000\times1000$ are particularly well suited for the test.
The subextensive part of their block entropies diverges \cite{nos2,nos1}, 
implying the presence of very long-range correlations 
that make the entropy estimation particularly difficult.
For these symbolic sequences we can not obtain the 
exact entropy values with analytical methods. Instead, we can use 
a reliable block-entropies method   \cite{Feldman} which estimates 
the entropy 
by using differential entropies ($S_B$).
Results are robust, as alternative block-entropies methods give 
equivalent outcomes \cite{nos2}.

\begin{figure}[t]
\begin{center}
\vspace{5mm}
\includegraphics[width=0.45\textwidth, angle=0]{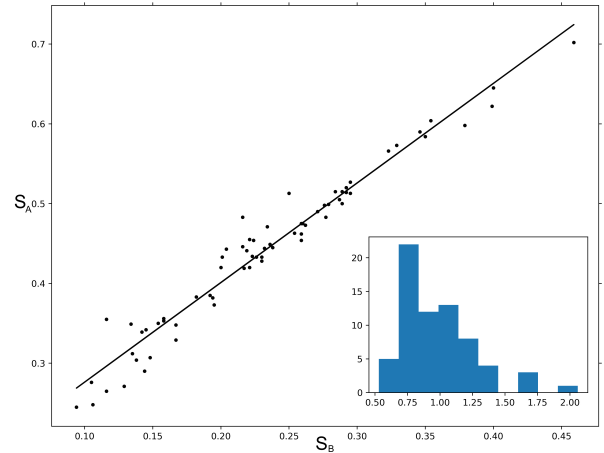}
\end{center}
\caption{\small  Scatter plot of the 
$S_A$ and $S_B$ values. 
$S_A$ are estimated by using the method in \cite{Avinery} with the Hilbert's curve. 
$S_B$ are obtained following the scheme of equation 3 in \cite{nos2}, 
for blocks of size up to $15$.
The continuous line represents the linear fitting: $y=1.25x+0.15$.
In the inset, the distribution of ($S_A$-$S_B$)/$S_B$. 
}
\label{fig_fragments}
\end{figure}

Fig.1 compares the results obtained by the classical 
block-entropies method ($S_B$) with the ones generated by the 
compression algorithm 
($S_A$).
This second approach dramatically overestimates the entropy.
The median of the  percentage error is 91\% and the scattering
plot shows that the difference between $S_A$ 
and $S_B$ grows for larger $S_B$ values.
The reduction of 2D patterns to a 1D string
significantly destroys the involved bidimensional structures,
which are particularly significant in systems with 
long-range correlations, and
generates a substantial overestimation of the entropy.

Even if Avinery's 
method is indeed elegant 
and computationally effective, 
it can not be considered accurate for general 2D systems. \\

E. Brigatti
\begin{quote}
Instituto de F\'{\i}sica, Universidade Federal do Rio de Janeiro, 
Av. Athos da Silveira Ramos, 149,
21941-972, Rio de Janeiro, RJ, Brasil.
\end{quote}

F.N.M. de Sousa Filho
\begin{quote}
Instituto de Computa\c c\~ao, Universidade Federal do Rio de Janeiro,
Av. Athos da Silveira Ramos, 274,
21941-916, Rio de Janeiro, RJ, Brasil.
\end{quote}



\end{document}